\documentclass[manuscript]{aastex}

\usepackage{graphicx}
\usepackage{rotating}
\usepackage{dcolumn}
\usepackage{natbib}
\usepackage{longtable}

\newcommand{\rsun}{R_{\Sun}}
\newcommand{\rstar}{R_{*}}
\newcommand{\mydeg}{^{\circ}}

\shorttitle{dM CME Deflections}
\shortauthors{Kay et al.}

\begin{document}

\title{Probability of CME Impact on Exoplanets Orbiting M Dwarfs and Solar-Like Stars}

\author{C. Kay}
\affil{Solar Physics Laboratory, NASA Goddard Space Flight Center, Greenbelt, MD, 20771}
\affil{Astronomy Department, Boston University, Boston, MA 02215}
\email{ckay@bu.edu}

\author{M. Opher}
\affil{Astronomy Department, Boston University, Boston, MA 02215}

\and 

\author{M. Kornbleuth}
\affil{Astronomy Department, Boston University, Boston, MA 02215}

\begin{abstract}
Solar coronal mass ejections (CMEs) produce adverse space weather effects at Earth.  Planets in the close habitable zone of magnetically active M dwarfs may experience more extreme space weather than at Earth, including frequent CME impacts leading to atmospheric erosion and leaving the surface exposed to extreme flare activity.  Similar erosion may occur for hot Jupiters with close orbits around solar-like stars.  We have developed a model, Forecasting a CME's Altered Trajectory (ForeCAT), which predicts a CME's deflection.  We adapt ForeCAT to simulate CME deflections for the mid-type M dwarf V374 Peg and hot Jupiters with solar-type hosts.  V374 Peg's strong magnetic fields can trap CMEs at the M dwarfs's Astrospheric Current Sheet, the location of the minimum in the background magnetic field.  Solar-type CMEs behave similarly, but have much smaller deflections and do not get trapped at the Astrospheric Current Sheet.  The probability of planetary impact decreases with increasing inclination of the planetary orbit with respect to the Astrospheric Current Sheet - 0.5 to 5 CME impacts per day for M dwarf exoplanets, 0.05 to 0.5 CME impacts per day for solar-type hot Jupiters.  We determine the minimum planetary magnetic field necessary to shield a planet's atmosphere from the CME impacts. M dwarf exoplanets require values between tens and hundreds of Gauss.  Hot Jupiters around a solar-type star, however, require a more reasonable $<$30 G.  These values exceed the magnitude required to shield a planet from the stellar wind, suggesting CMEs may be the key driver of atmospheric losses.
\end{abstract}

\keywords{Sun: coronal mass ejections (CMEs)}

\section{Introduction}
Recent observations by the Mars Atmosphere and Volatile Evolution (MAVEN) spacecraft have shown that coronal mass ejections (CMEs) can have a significant impact on the Martian atmosphere, and may have influence Mars' long-term atmospheric evolution \citep{Jak15}.  Increases in the external ram and magnetic pressure, due to the passage of the CME, compress the Martian magnetosphere, which in turn affects the ionosphere and neutral atmosphere.  Comparison of MAVEN observations and numerical modeling show that the impact of a CME on 08 March 2015 caused over an order of magnitude increase in the ion escape rate on the dayside, as compared to quiescent times.

Evidence from our own solar system suggests that space weather can affect the habitability of a planet through atmospheric losses \citep{Jak15}.  For other stellar systems, we expect space weather to be important.  Additionally, extrasolar space weather may differ significantly from our own system as exoplanets can orbit at very close distances, and the stellar activity and magnetic field strength can be greatly enhanced.  In this work we combine knowledge of solar space weather with inferred properties of other systems to determine the frequency and severity of CME impacts for two specific cases - a hot Jupiter orbiting a solar-type star, and an exoplanet in the habitable zone of a M dwarf.

\subsection{Hot Jupiters}
``Hot Jupiters'' were originally one of the most frequently discovered types of exoplanet.  Hot Jupiters tend to orbit solar-like stars (F-, G-, and K-type) at very close distances, typically of order 10 stellar radii ($\rstar$, 0.046 AU) and as small as 3 $\rstar$ \citep{Heb09}.  The close orbits lead to a systematic bias in the frequency of observed hot Jupiters - these planets are the easiest to observe using radial-velocity or transits.  For example, the transit of the exoplanet HD189733b results in a photometric depth of 3\% in the light curve \citep{Bou05}.  While hot Jupiters may account for approximately 20\% of discovered exoplanets, \citet{Wri12} suggest that only 1\% of F, G, and K type stars host an hot Jupiter.

Hot Jupiters are not habitable exoplanets due to their close orbits and large size.  However, hot Jupiters present an opportunity to study planetary systems unlike anything else in our own solar system.  The study of hot Jupiters has yielded new insights on the evolution of planetary systems as hot Jupiters are thought to form at farther radial distances and migrate in toward the star (e.g. \citet{Koz62} and \citet{Lin96}).  Hot Jupiters are expected to have very different planetary weather than seen in the solar system.  The large asymmetry between the dayside and nightside temperatures can lead to extreme atmospheric winds as fast as 3000 m s$^{-1}$ \citep{Kat13}.

The small orbital distances lead to high levels of insolation, which can inflate the radii of hot Jupiters and lead to lower planetary densities \citep{Bur07}.  The high levels of XUV radiation can cause the inflated atmospheres to escape at extreme rates.  For HD189733b, \citet{Vid03} determine a minimum atmospheric escape rate of 10$^{10}$ g s$^{-1}$, with observations consistent with values up to several orders of magnitude higher.  More recent measurements by \citet{Lec10} are consistent with a rate of 10$^{10}$ g s$^{-1}$.  The hot, inflated atmosphere would be extremely vulnerable to erosion from CME impacts unless shielded by a planetary magnetic field \citep{Lam06, Kho07HJ}. For comparison, \citet{Jak15} estimate that the 08 March 2015 CME impact increased the Martian atmospheric escape to 10$^{4}$ g s$^{-1}$.  The quiescent atmospheric loss of a hot Jupiter greatly exceeds values seen in the solar system, and could be significantly enhanced by CME impact. 

\subsection{M Dwarfs}
Small, low mass stars, such as M dwarfs, vastly outnumber their more massive counterparts within the Galaxy.  M dwarfs' great number, combined with their long main sequence lifetime, and small mass and radius, which results in a low luminosity, have made them popular targets in the search for habitable exoplanets. 

However, many of the aspects that make M dwarf exoplanets easier to find may also be a detriment to their habitability.  The cool, low luminosity nature of M dwarfs leads to close habitable zones, as defined using the traditional requirement of planetary temperatures conducive to the existence of liquid water.   M dwarf habitable zones range between 0.03 Astronomical Units (AU) and 0.4 AU, with the distance being the smallest for late-type M dwarfs \citep{Kho07}.  The largest of these orbital distances is roughly the same as Mercury's orbit.  For the early- and mid-type habitable zone exoplanets, we have no analogue orbit at such small distances within our own solar system.   

Stellar activity tends to increase with the size of the stellar convection envelope \citep{Wes04} and stellar rotation rates \citep{Moh02,Wes15}, although the activity saturates for sufficiently high rotational velocity \citep{Def98}.  For mid- to late-type M dwarfs (M4 to M8.5) the activity saturates at higher rotational velocities than for early-type M dwarfs, and above M9 the activity levels decrease significantly \citep{Moh02}.  Accordingly, most M dwarf stars will have significantly enhanced stellar activity as compared to the Sun.  The effects of this enhanced activity should be more pronounced relative to solar exoplanets due to the close proximity of a ``habitable'' M dwarf exoplanet.  M dwarfs have extremely long main sequence lifetimes and can remain active for periods of order Gyrs \citep{Wes04} potentially impinging on exoplanet habitability for long time scales.

One manner in which space weather can adversely affect a potentially habit exoplanet is via the stellar X-ray and EUV flux (XUV).  This radiation can heat the upper atmosphere or even ionize it, leading to atmospheric ion pick up loss \citep{Lam07}. If the UV radiation can penetrate to the surface then it can damage any potential DNA present \citep{Sca07}.  While the quiescent XUV flux for a M dwarf habitable zone planet is likely an order of magnitude less than that at Earth, the flux will increase to 10-100 times that at Earth during M dwarf flares \citep{Sca07}.

\citet{Sca07} suggest that retaining a moderate atmosphere is critical for habitability and that this can be facilitated by the presence of a strong planetary magnetic field.  However, \citet{Gri05,Gri09} suggest that planets orbiting M dwarfs at distances less than 0.2 AU may be tidally locked and the slow rotation will affect the planetary dynamo, leading to little to no planetary magnetic field.  \citet{Kho07} show that in the case of a significant planetary magnetic field, the atmosphere can still be eroded when CME impacts compress the planetary magnetosphere.

\subsection{CME Impacts}
As the number of CME impacts increases, an exoplanet becomes less likely to retain an atmosphere to shield the surface from harmful XUV radiation.  This can adversely affect the habitability of an M dwarf exoplanet in the habitable zone, or potentially erode the hot Jupiter itself.  These effects will be minimized for planets with an orbital plane corresponding to locations where CMEs are less likely to impact.  Recent observations of M dwarfs indicate that multi-planetary systems are common, and that the planets tend to share an orbital plane \citep{Lis11, Tre12, Fan12, Fab14, Bal15, Cro15}.  For a planet orbiting a star which formed from the same gaseous disk we expect low obliquity (angle between the planetary orbit and the stellar rotation), however, many counterexamples with high obliquity have been observed \citep{San12, San13, Bou14, Daw14}. \citet{Mor14} find that the orbits of single planets tend to be more oblique than multi-planet systems.  In this work we assume low obliquity, which is representative of many planetary systems, but certainly not all.

After assuming some planetary orbit, to be able to predict the likelihood of CME impact, we need to understand the path a CME takes as it leaves a star. Observations show that solar CMEs ``deflect,'' or deviate from a purely radial trajectory.  These deflections correspond to changes in latitude, longitude, or both \citep{Kil09, Isa13, Rod11, Wan02, Wan04, Wan06, Wan14}.  The deflections tend to move CMEs away from coronal holes (CHs) toward the Heliospheric Current Sheet (HCS, \citet{Cre06,Kil09, Gop09, Moh12}).  The HCS corresponds to where the solar magnetic field reverses radial polarity, and tends to be located near the solar equator during solar minimum conditions.  Magnetic forces deflecting CMEs toward the minimum magnetic energy, the HCS on global scales, can explain the direction of observed CME deflections \citep{Gui11, She11, Kay13, Kay15L, Kay15}.

In this paper, we adapt a solar CME deflection model, Forecasting a CME's Altered Trajectory (ForeCAT; \citet{Kay13,Kay15}), to predict the deflections of CMEs for an M dwarf and hot Jupiter system.   Considering the full range of plausible CME masses and velocities for both cases, we determine whether CME deflections can increase the likelihood of exoplanetary impacts, which may be detrimental to M dwarf exoplanet habitability or increase the atmospheric erosion of hot Jupiters.

\section{ForeCAT}
\citet{Kay13,Kay15} developed ForeCAT, a model for solar CME deflections and rotations based on the background solar magnetic forces (magnetic tension and magnetic pressure gradients).  ForeCAT neglects some time-varying effects such as reconnection.  In contrast to more sophisticated magnetohydrodynamic (MHD) models, ForeCAT includes the minimum physics necessary to accurately reproduce observed solar CMEs \citep{Kay15L, Pis15} making large parameter space studies feasible on a standard GPU-enabled desktop computer.

ForeCAT CMEs are initiated by specifying the initial position of a CME on the stellar surface (latitude, longitude, and tilt, measured clockwise with respect to the equatorial plane), three shape parameters to describe an elliptical toroidal shape \citep{Kay15}, as well as the CME mass, $M$, and final propagation velocity, $v_f$.  ForeCAT includes drag in the nonradial direction to slow the deflection motion.  The standard hydrodynamic drag equation can describe MHD drag \citep{Car96, Car04} and works best with a drag coefficient set equal to a constant times the hyperbolic tangent of the plasma $\beta$ (the ratio of thermal to magnetic pressure).  The drag coefficient constant is taken to be one throughout this work.  \citet{Kay15} and \citet{Pis15} show that ForeCAT reproduces the observed trajectory of solar CMEs and is able to constrain some of the CMEs initial parameters and the lower corona background.

ForeCAT depends on empirical and analytic models to describe a CME's radial propagation and expansion.  The CME follows a three-phase propagation model, similar to that presented in \citet{Zha06}- a slow rise followed by rapid acceleration and finally constant radial propagation. ForeCAT uses fixed radial distances to transition between the three phases.  The CME slowly rises at 80 km s$^{-1}$ until 1.5 $\rsun$, followed by linear acceleration until it reaches $v_f$ at 3 $\rsun$.  \citet{Kay15} explore the effect of these chosen values, which are reasonable average values for solar CMEs, and find that variations in these parameters can have small effects on the magnitude of the deflection.  The global trends in CME deflection, however, do not change, and we expect that these chosen values should not significantly effect this work.  These effects will be explored in a future work.  We expect that the higher surface magnetic field strength of a M dwarf \citep{Mor10} will provide more energy to accelerate CMEs radially.  This could cause an increase in the initial velocity if the CMEs have comparable masses to solar CMEs, as discussed in Section \ref{dMparams}.  While individual deflections are sensitive to the chosen parameters of the three-phase propagation model \citep{Kay15}, we find that the global trends of M dwarf CME deflections do not depend significantly on their specific values.  

All ForeCAT CMEs expand self-similarly in this work, maintaining a fixed angular width. Observations of solar CMEs suggest that their angular width, as viewed in a coronagraph, can rapidly increase close to the Sun but remains constant beyond approximately 5 $\rsun$ \citep{Che00,Cre04,Pat10a,Pat10b}.  However, self-similar expansion has also been observed much closer to the Sun \citep{The06,Asc09,Woo09}.

The Potential Field Source Surface (PFSS) model describes the solar background magnetic field using a sum of Legendre polynomials weighted by harmonic coefficients determined from a photospheric magnetogram. This model assumes that the magnetic field can be described as current-free and becomes entirely radial above the ``source surface.''  The source surface is typically taken to be 2.5 $\rsun$ for PFSS models of the Sun as this produces the most accurate representation of the global structure of the magnetic field \citep{Hoe84}.  The PFSS model has recently been used to describe the magnetic field of a M dwarf \citep{Don06} in stellar wind simulations \citep{Vid11,Vid14}. The appropriate source surface distance for an M dwarf is unknown, but \citet{Vid11,Vid14} use values of 5 or 4 $\rstar$, respectively.  Above the source surface, the solar rotation causes the magnetic field to follow an Archimedian spiral \citep{Par58}.  

The solar wind density is determined by a modified version of the (\citet{Guh06}, hereafter G06) density model. The G06 density model combines empirical relations describing the polar or current sheet density versus radial distance.  The radial distance from the current sheet determines the relative contribution of the two density profiles.  Finally, we assume the solar wind is purely radial so that the velocity can be determined using the density and a mass flux that does not vary with distance.  Similar to the G06 model, we allow the mass flux to vary with angular distance from the current sheet. We discuss the adaptation of this model to a M dwarf stellar wind in Section \ref{stellarwind}.

\section{Stellar Wind Parameters}\label{stellarwind}
V374 Peg is an M4 star with radius $R_*=$ 0.34 $\rsun$, mass $M_*=$ 0.28 $M_{\sun}$, and a 0.44 day rotation period.  The habitable zone, as defined using the incident stellar flux \citep{Kop13}, should exist around 0.1 AU (20 $\rsun$,  60$\rstar$).  While smaller mass M dwarfs will have even closer habitable zones that may lead to more extreme space weather, we use V374 Peg in this work as it has a well-studied surface magnetic map, reconstructed through the use of Zeeman Doppler Imaging \citep{Don06}.  This map of the magnetic field has been used to drive several MHD simulations of the stellar wind \citep{Vid11, Kor15}.  We use the V374 Peg magnetogram and a source surface radius of 5 $R_*$, the same as used in \citet{Vid11} and \citet{Kor15}, to determine the magnetic field with the PFSS model.  To study the effects on hot Jupiters orbiting solar-type stars, we simulate a solar-type star with a magnetic background corresponding to Carrington Rotation (CR) 2029 (2005 April-May), used previously with ForeCAT in \citet{Kay15}. In the results of this paper, when referring to trends for ``M dwarfs'' versus ``hot Jupiters'' we specifically mean mid-type M dwarfs and hot Jupiters around solar-type stars.  We expect the results to change for different spectral types. 

\subsection{Stellar Wind Density}
For the background stellar wind, we use the results of the MHD solutions presented in \citet{Kor15}.  \citet{Kor15} consider two types of solar wind heating: a non-isothermal-driven wind where the wind is heated by a spatially-varying polytropic index \citep{Coh07}, and an Alfv\'{e}n-driven model where the wind is accelerated and heated by Alfv\'{e}n waves that are damped by surface Alfv\'{e}n waves damp and/or turbulence \citep{Van10, Eva12}.

For V374 Peg we use the Alfv\'{e}n and non-isothermal driven models \citep{Kor15}, which we refer to as V374 ALF and V374 TER in this work.  For CR 2029 we consider one Alfv\'{e}n case and two non-isothermal cases, referred to as CR2029 ALF and CR2029 TER1 and CR2029 TER2 in this work.  CR2029 TER1 uses the same unscaled magnetic field and coronal boundary density and temperature as in CR2029 ALF.  For CR2029 TER2, the background magnetic field is scaled up by a factor of 4 in the MHD simulation, a common practice with the non-isothermal driven heating.  Additionally, CR2029 TER2 has a larger coronal base density and hotter coronal temperature than CR2029 ALF.  

\begin{figure}[!hbtp]
\includegraphics[height=5in, angle=0]{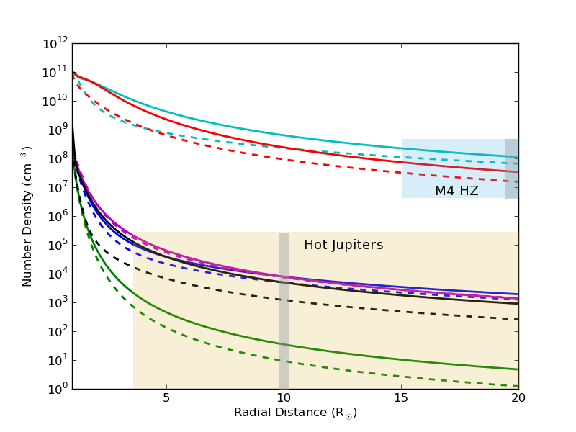}
\caption{The number density versus radial distance for the V374 ALF (cyan), V374 TER (red), CR2029 ALF (blue), CR2029 TER (green), CR2029 TER2 (magenta), and G06 (black).  The solid and dashed lines correspond to the current sheet and polar profiles of each model.  The blue and orange shaded regions correspond to the distances of a M4 dwarf habitable zone or hot Jupiters.  The grey shaded region corresponds to the distance used in this work.}\label{fig:dens}
\end{figure}

For each case we determine the coefficients for the density model of \citet{Guh06} (hereafter G06) by fitting to the MHD models. The G06 density model determines the solar wind density using the combination of a current sheet and a polar radial density profile,
\begin{equation}\label{eq:G06b}
\rho_x = a_1 \mathrm{e}^{a_2 z + a_6 z^2}z^2[1+a_3 z + a_4 z^2 + a_5 z^3] \mathrm{,}
\end{equation}
where the $x$ represents either the current sheet (CS) or the polar (P) values.  The relative contribution of each polynomial is weighted by $\lambda$, the distance from the current sheet,
\begin{equation}\label{eq:G06a}
\rho(R, \theta) = \rho_p(R) + [\rho_{cs}(R) - \rho_p(R)] \mathrm{e}^{-\lambda^2 / w^2} \mathrm{,}
\end{equation}
where $w$ is a measure of the angular width of the current sheet.  We fit the width from the MHD solution using a second order polynomial below 4.5 $\rsun$ or 4.5 $\rstar$
\begin{equation}\label{eq:CSw}
w = w_1 + w_2 R + w_3 R^2
\end{equation}
where the radial distance $R$ has units of $\rsun$ or $\rstar$ for the solar and M dwarf cases, respectively.  Figure \ref{fig:dens} shows the number density versus radial distance for the five density models used in this work, as well as the G06 model.  For all cases the current sheet density (solid lines) exceeds the polar density (dashed lines).  V374 ALF and V374 TER (cyan and red lines, respectively) behave similarly close to the star, but V374 ALF exceeds V374 TER at farther distances.  Both M dwarf models exceed the solar models by several orders of magnitude at all distances.   Both the current sheet and polar components of CR2029 ALF and CR2029 TER2 (blue and magneta, respectively) behave similarly to the G06 current sheet profile (solid black line).  CR2029 TER is orders of magnitude smaller than the other solar models, particularly at large distances.

\subsection{Stellar Wind Speed}
We determine the stellar wind speed using the density and the assumption of a constant mass flux along a radial path.  The mass flux values are the same as those given in \citet{Kor15}.  As done for previous solar ForeCAT results, we use a higher mass flux in the poles than the current sheet and determine the value at any given location using a weighting based on angular distance from the current sheet.  CR2029 TER1 produces mass fluxes significantly less than observed solar values.  We find that ForeCAT is not terribly sensitive to the chosen values, their effects only become significant at distances at which the deflection has become negligible.

\section{CME Parameters}\label{dMparams}
Coronagraph images have been instrumental to our current understanding of solar CMEs. A CME's plane-of-the-sky velocity can be measured as the time-derivative of its radial position and the increase in brightness relative to the solar background provides an estimate of the CME mass \citep{Vou00}.  

Directly imaging extrasolar CME's remains implausible for the foreseeable future so the properties of stellar CMEs must be inferred by other means.  The signatures of stellar CMEs have been searched for via type II radio bursts \citep{Jac90, Abd95, Abr98}, X-ray dimmings \citep{Jen86}, UV absorption \citep{Sch83}, and more recently, an enhancement in the blue wing of a spectral line due to the Doppler shift of a propagating CME \citep{Hou90, Den93, Din03, Gue97, Fuh04, Lei14}.  Successful observations of blue-wing enhancements have yielded CME velocities between a few hundred to many thousands of km s$^{-1}$.  These observations suggest stellar CMEs may have masses and velocities greatly exceeding the average solar values of a few times 10$^{14}$ g and approximately 500 km s$^{-1}$ \citep{Gop09LASCO}.  \citet{Hou90} infer an 8x10$^{17}$ g CME erupting at 5800 km s$^{-1}$ from the young M3.5 dwarf AD Leo and \citet{Gue97} estimate a 10$^{18}$ to 10$^{19}$ g CME from a T-Tauri star.  \citet{Vid16} recently infer a CME with mass greater than 10$^{16}$ g from V374 Peg.

The plausible range of M dwarf CME parameters can be approximated based on observations of M dwarf flares combined with scaling relations between CMEs and flares determined from solar measurements.  \citet{Aar11,Aar12} determine an empirical relationship between solar flare energy and CME mass and show that it can be extrapolated to other stellar types with larger flare energies.  \citet{Aud00} analyze the flare rates of late-type stars (spectral type F-M) and determine that, on average, a few 10$^{32}$ erg flares occur per day.  This flare energy is equivalent to that of the largest observed solar flares \citep{Sch12}.  Using the flare energy-CME mass relation from \citet{Aar12}, we expect a M dwarf to release several 10$^{17}$ g CMEs per day. Recent observations of V374 Peg showed flare rates between 0.15 and 0.42 flares per hour, with maximum total energies as large as 10$^{33}$ ergs \citep{Vid16}.

To determine the plausible range of M dwarf CME speeds, we first consider the range of solar CME speeds.  It is generally assumed that some fraction of an active region's free magnetic energy is converted into a CME's kinetic energy (or the quiet sun magnetic energy in the case of filament eruptions).  \citet{Ven03} and \citet{Che06} find a strong correlation between CME speed and the active region magnetic energy.  Using the data of \citet{Ven03} and making the approximation of a constant mass for all CMEs, we find between 0.5\% and 5\% of the available magnetic energy tends to be converted into kinetic energy depending on the chosen CME mass.  \citet{Gop04} consider the available energy in a large active region (diameter 5 arcmin, total volume $V$=10$^{30}$ cm$^3$) with average photospheric magnetic field strength 200 G.  This yields an available energy of 1.6x10$^{33}$ ergs.  For a more average size AR (1000 Mm$^2$ $\approx$ 0.8 arcmin diameter, \citet{How96}) we find an available energy of 2.5x10$^{31}$ ergs.  We determine a typical CME mass by assuming a half-torus of major and minor radius 0.2 and 0.05 $\rsun$ filled with the coronal base density (8.35x10$^{-16}$ g cm$^{-3}$, \citet{Guh06}) which yields a mass, $M$, of 1.4x10$^{15}$ g which is slightly larger, but similar to the average observed CME mass of 1.3x10$^{15}$ g \citep{Vou10, Vou11er}.  Assuming some fraction $\alpha$ of the total magnetic energy becomes kinetic energy, we can determine the CME speed, $v_{CME}$.
\begin{equation}\label{eq:vmax}
v_{CME} = \sqrt{\frac{\alpha B^2 V}{4 \pi M}}
\end{equation}
For $\alpha$=0.05 and using the large active region volume, we find a maximum speed of approximately 3400 km s$^{-1}$, and for the average active region volume, we find a speed of approximately 425 km s$^{-1}$.  These speeds are in good agreement with the maximum and average observed solar CME velocities.

We now apply this estimation, shown to be reasonable for solar CMEs, to mid-type M dwarf CMEs.  We assume the active region and CME retain the same size relative to their host star.  The M dwarf volumes correspond to $(\rstar / \rsun)^3$, or 3.9\%, of their solar values due to the smaller radius of V374 Peg ($\rstar$=0.34$\rsun$).  Our M dwarf density model has a coronal base density of 1.67x10$^{-13}$ g cm$^{-3}$, which yields a CME mass of 1.66x10$^{16}$ g, about an order of magnitude larger the average solar CME mass.  Observations of M dwarf surface magnetic field strength tend to lack sufficient resolution to resolve individual active regions.  The average active region B of 200 G used in the previous calculation is approximately 20 times the average solar quiet sun photospheric value, so we scale the observed M dwarf magnetic field of ~1 kG to 20 kG in a M dwarf active region.  Using Equation \ref{eq:vmax} with these values yields an average speed of 4,013 km s$^{-1}$ for a 1.66x10$^{16}$ g CME, and a maximum speed of 31,600 km s$^{-1}$.  

To cover a reasonable range of the plausible M dwarf parameter space, we simulate CMEs between 10$^{14}$ g to 10$^{19}$ g.  The lower limit represents the least massive solar CMEs relevant for space weather and the upper limit corresponds to the maximum stellar values inferred from observations.  We consider M dwarf speeds from 300 km s $^{-1}$ up to 10,000 km s$^{-1}$.  For the more massive CMEs we restrict the speed to the upper limit, $v_{max}$, determined using Eq. \ref{eq:vmax} and the large M dwarf active region volume (3.9x10$^{28}$ cm$^3$) with an average B of 20 kG.
\begin{equation}\label{vmax2}
v_{max} = 31600 \left( \frac{M}{\mathrm{10^{15} g}}\right)^{-1/2} \mathrm{km \; s^{-1}}
\end{equation}
For the solar-type simulations we use the observed solar CME ranges of masses between 10$^{14}$ g to 10$^{16}$ g and speeds between 300 km s $^{-1}$ and 1,500 km s$^{-1}$.

\section{ForeCAT Results for V374 Peg CME Deflections}
Using ForeCAT, we simulate the deflections of CMEs out to 0.1 AU ($~$60 $R_*$).  All CMEs begin at a longitude of 150$\mydeg$, and we consider high, mid, and low latitude CMEs, which respectively start at 70$\mydeg$, 40$\mydeg$, and the equator.  The CMEs are oriented parallel to the equator and we do not consider any effects of rotation in this work. For each initial position we initiate CMEs with masses between 10$^{14}$ g to 10$^{19}$ g and speeds between 300 and 10,000 km s$^{-1}$, or the maximum speed determined by the mass and Equation \ref{eq:vmax}.

\subsection{Individual Cases}
\begin{figure}[!hbtp]
\includegraphics[width=5in, angle=0]{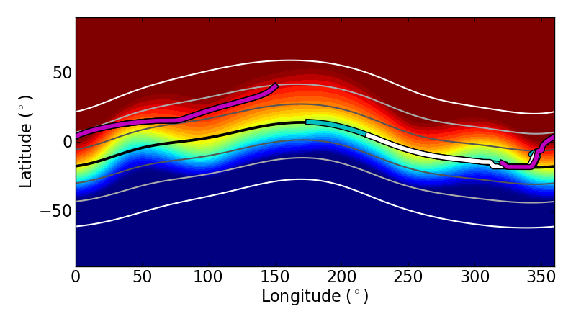}
\caption{The effect of the final propagation speed on the deflection of a CME.  The color and line contours represent the background magnetic field, as described in the text, and the lines represent the trajectory out to 0.1 AU for a 300 km s$^{-1}$ CME (cyan), a 1,000 km s$^{-1}$ CME (white), and a 5,000 km s$^{-1}$ CME (purple). A shift in longitude due to stellar rotation is incorporated into the CME trajectory.}\label{fig:vels}
\end{figure}

Before looking at results for the full mass and speed parameter space, we consider a few individual cases to better understand the differences between solar and M dwarf CME deflection.  We emphasize that in this work we are looking at a single mid-type M dwarf, and that the behavior may differ greatly for other types, which we discuss in Section \ref{disc}. Figure \ref{fig:vels} shows the trajectory of mid-latitude CMEs with masses of 10$^{14}$ g and speeds of 300 km s$^{-1}$ (cyan), 1,000 km s$^{-1}$ (white), and 5,000 km s$^{-1}$ (purple).  Note that the cyan curve for the lowest mass case barely extends past the white curve. In each panel the color contours show the radial magnetic field and the line contours show the magnetic field strength at the source surface.  The black contour line gives the approximate location of the current sheet, the minimum in the magnetic energy at farther distances.  We refer to the stellar analogue of the Heliospheric Current Sheet as the Astrospheric Current Sheet.  Based on solar CME deflections, we expect stellar CMEs to deflect to the Astrospheric Current Sheet.

Stellar rotation causes the background to change with respect to a CMEs position, even if the CME is not deflecting.  CMEs wtih different final propagation speeds will experience different amounts of stellar rotation so rather than shifting the background, we add the translation due to rotation into the CMEs' trajectory in Figure \ref{fig:vels}.  A CME that does not deflect would appear as a line of constant latitude in Figure \ref{fig:vels}.

The CMEs initially behave the same, deflecting eastward and toward lower latitudes, as they have the same speed during the slow rise phase.  In Figure \ref{fig:vels} the trajectories overlap until the acceleration phase begins.  All three CMEs move toward the Astrospheric Current Sheet and upon reaching it they become ``trapped'' as they cannot penetrate the potential barrier on the opposite side.  This is evident in Figure \ref{fig:vels} as the combined rotation and deflection motion clearly traces the path of the Astrospheric Current Sheet.  

For all cases, as stellar rotation changes the CME's stellar longitude changes, the latitude of the Astrospheric Current Sheet changes, causing the CME to deflect latitudinally.  The stellar magnetic field is strong enough to force low mass CMEs to continue sliding along the ACS for the duration of the propagation out to 0.1 AU.  Faster CMEs experience less change in longitude due to the decrease in propagation time.  This behavior is more extreme than what we have previously seen for solar CME deflection \citep{Kay15}.  While solar CMEs deflect toward the Heliospheric Current Sheet, the solar magnetic forces are not strong enough to produce this trapping behavior, even for the less massive solar CMEs.  

\begin{figure}[!hbtp]
\includegraphics[height=7in, angle=0]{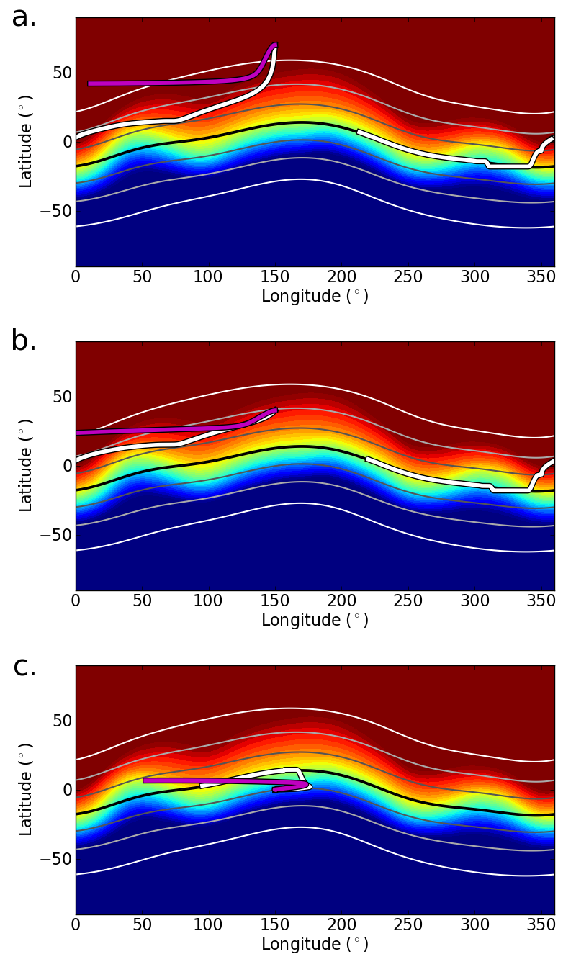}
\caption{The effect of initial position and CME mass on the CME deflection, in the same format as Figure \ref{fig:vels}.  The effects of stellar rotation have been incorporated into the CME trajectory.  The top, middle, and bottom panels correspond to high, mid, and low initial CME latitudes.  The white lines correspond to 10$^{14}$ g CMEs and the purple lines to 10$^{18}$ g CMEs.}\label{fig:MassPos}
\end{figure}

We also consider the effects of different CME masses and initial starting locations.  Figure \ref{fig:MassPos} shows results in the same format as Figure \ref{fig:vels} but for CMEs with a speed of 1000 km s$^{-1}$.  Analogous to Figure \ref{fig:vels}, the effects of rotation have been incorporated into the CME trajectory.  The top, middle, and bottom panels correspond to the high, mid, and low initial latitudes, respectively, and the white trajectories correspond to a 10$^{14}$ g CME and the purple trajectories to a 10$^{18}$ g CME.

The white line in the middle panel of Figure \ref{fig:MassPos} corresponds to the same case as the white line in Figure \ref{fig:vels}.  The top and bottom panels show that the low mass CMEs behave similarly, regardless of their initial location.  Although the CMEs deflect to the Astrospherical Current Sheet, the actual path taken will vary.  Upon reaching the Astrospherical Current Sheet the CMEs remain trapped, experiencing very little longitudinal motion in the inertial frame.  The effect of stellar rotation on the position of the Astrospherical Current Sheet then forces the CME to change latitude. 

As seen in \citet{Kay15} the deflection decreases with CME mass, however, significant deflections ($>$20$\mydeg$) can still occur for 10$^{18}$ g CMEs.  For these high mass CMEs, the deflection tends to cease by 5 $\rstar$.  In the CME trajectories in Figure \ref{fig:MassPos} this corresponds to a horizontal line toward the east.  When the Astrospheric Current Sheet is inclined with respect to the stellar equator, the motion due to rotation can move the CME away from the magnetic minimum.  Unlike for the low mass CMEs, the magnetic forces are not strong enough to cause the high mass CMEs to slide along the Astrospheric Current Sheet.

\subsection{Full Parameter Space}

\begin{figure}[!hbtp]
\includegraphics[height=6in, angle=0]{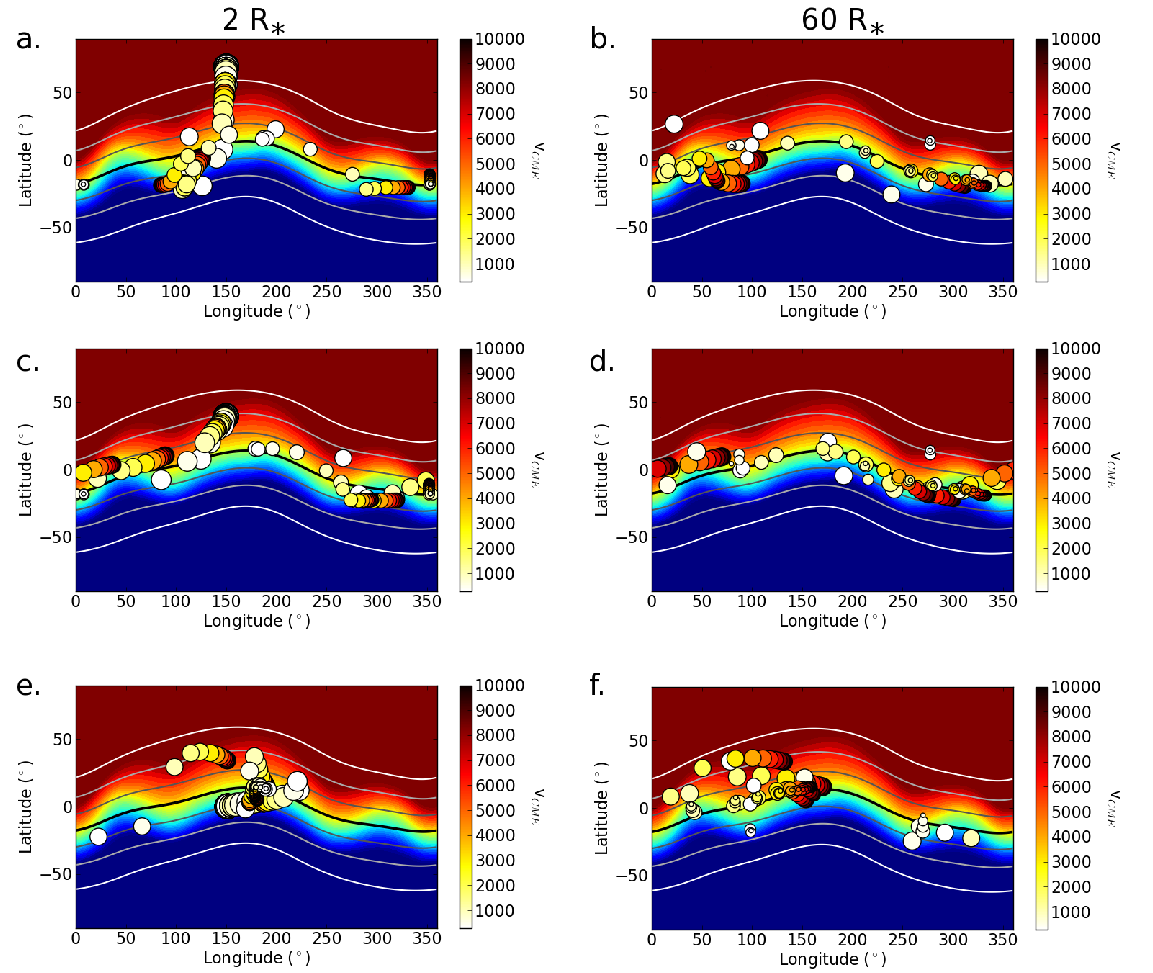}
\caption{Deflected positions of CMEs sampling mass and velocity parameter space for the M dwarf V374 Peg.  The left and right panels show the positions at 2 $R_*$ and 0.1 AU (60 $R_*$), respectively.  Each circle represents a CME and the size and color indicate the CME mass and velocity.  The color and line contours represent the background magnetic field as in Figure \ref{fig:vels}.  The effects of stellar rotation have been incorporated into the CME positions.}\label{fig:MvV}
\end{figure}

Figure \ref{fig:MvV} shows the results of 115 simulations for each initial latitude spanning a range of CME masses and final propagation speeds.  The color and line contours represent the magnetic background as in Figures \ref{fig:vels} and \ref{fig:MassPos}. Each circle represents a single CME, the size of the circle indicates the CME mass with more massive CMEs being larger.  The color of each circle represents the CME's final speed.  The left panels show the position of each CME at 2 $R_*$ and the right panels at the habitable zone distance, 0.1 AU.  As in Figures \ref{fig:vels} and \ref{fig:MassPos} we have incorporated a change in longitude due to stellar rotation to better show the position of the CMEs with respect to the appropriate portion of the Astrospheric Current Sheet.  The most massive CMEs, which have the smallest total deflections, show a negligible deflection between 2 $R_*$ and 0.1 AU, their motion in this range is predominantly due to the stellar rotation.  Accordingly, we only include CME masses as large as 10$^{17}$ g in the right panels of Figure \ref{fig:MvV}.  The top, middle, and bottom panels show results for the low, mid, and high latitude cases, respectively.  

The left panels of Figure \ref{fig:MvV} show a cluster of CMEs around the initial location corresponding to the most massive CMEs, which deflect the least.  For solar CMEs, the deflection is typically determined below 2 $\rsun$.  For the less massive CMEs, we find that while much deflection occurs below 2 $R_*$, a significant amount occurs at farther distances.  The lower mass CMEs are more tightly clustered around the Astrospherical Current Sheet at 0.1 AU than 2 $R_*$ for all initial latitudes.  For each initial latitude, we see that the CMEs with the same final speed tend to reach the same final longitude at 0.1 AU so their final longitude is determined by the propagation time.

\subsection{Current Sheet Distance}
ForeCAT results show that the strong magnetic fields of V374 Peg can cause significant deflections, which cause the majority of CMEs to move closer to the Astrospheric Current Sheet.  To understand the effect on exoplanet impact, we need to know the average distance of a M dwarf CME from the Astrospheric Current Sheet, which will depend on the CME mass.

\begin{figure}[!hbtp]
\includegraphics[height=4in, angle=0]{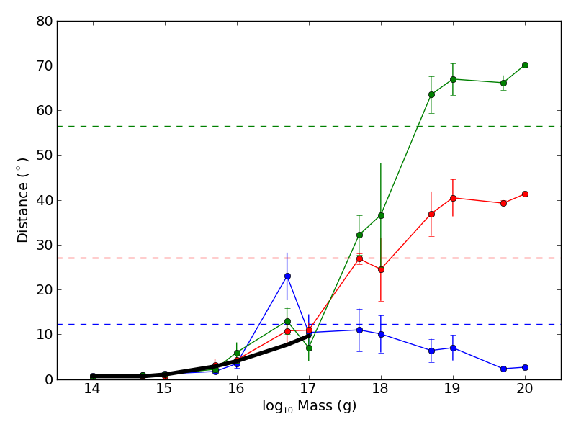}
\caption{The CME distance (in degrees) from the Astrospheric Current Sheet as a function of the CME mass.  For each initial CME latitude we determine the average distance of the CMEs from the Astrospheric Current Sheet at 0.1 AU (green:high, red:mid, blue:low).  The dashed lines show the initial distance from the Astrospheric Current Sheet.  The black line indicates the quadratic best fit to the results from all three initial latitudes for masses below 10$^{17}$ g.}\label{fig:CSdist}
\end{figure}

For each initial latitude we determine the average distance from the Astrospheric Current Sheet (including the effects of stellar rotation) as a function of CME mass.  Figure \ref{fig:CSdist} shows this quantity for the high (green), mid (red), and low (blue) initial latitudes.  The error bars correspond to one half of the standard deviation.  Small masses have small error bars because all the CMEs deflect close to the equator and show little scatter.  The error bars decrease for large masses because little to no deflection occurs for these CMEs.

Below approximately 5x10$^{17}$ g the distance from the Astrospheric Current Sheet decreases with mass for all cases.  Above this mass the trend is less clear and the distance depends more strongly on the initial distance (marked with a dashed line for each case).  We fit a quadratic polynomial to all three sets of data below 10$^{17}$ g to get a relation between CME mass, $M_{\mathrm{CME}}$, in $g$, and distance from the Astrospheric Current Sheet, $\Delta_{\mathrm{ACS}}$, in $\mydeg$, at 0.1 AU.
\begin{equation}\label{eq:deltaCS}
\Delta_{\mathrm{ACS}} =1.307 \log\left(M_{\mathrm{CME}}\right)^2 -37.53 \log\left(M_{\mathrm{CME}}\right) + 269.9
\end{equation}
The solid black line in Figure \ref{fig:CSdist} shows this fit.  Above 10$^{17}$ g we assume the CME deflection is negligible and CME remains at its initial distance from the Astrospheric Current Sheet.  The averages in Figure \ref{fig:CSdist} tend to be close to the initial distances for each mass.  As Figure \ref{fig:MassPos} shows, these massive CMEs do initially deflect toward the Astrospheric Current Sheet but after the deflection ceases the stellar rotation can cause an increase in the distance from the Astrospheric Current Sheet.

\section{ForeCAT Results for Hot Jupiter Distances}
\begin{figure}[!hbtp]
\includegraphics[width=5in, angle=0]{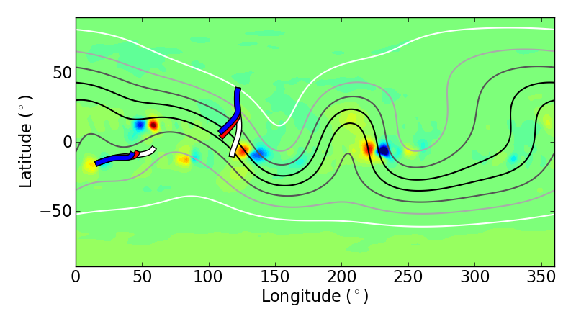}
\caption{Dependence of the CME deflections on the stellar wind density model for solar-type stars.  This background corresponds to the solar declining phase Carrington Rotation 2029.  The background color and line contours represent the magnetic field as in Figure \ref{fig:vels}. }\label{fig:HJdens}
\end{figure}

We repeat the analysis of the previous section for hot Jupiters orbiting solar-type stars at a distance of 10 $\rsun$ (0.05 AU), a distance representative of typical hot Jupiter orbits.  We simulate CMEs with masses between 10$^{13}$ g and 10$^{16}$ g and speeds between 300 and 1500 km s$^{-1}$, the same range seen for solar CMEs \citep{Vou10}.  Since we use a high resolution solar magnetogram we can resolve active regions and place the CMEs at their polarity inversion lines.  We consider one CME erupting from an active region (initial latitude and longitude of -15.4$\mydeg$ and 17$\mydeg$ with a tilt of -72$\mydeg$, hereafter active region CMEs) and one CME erupting from the quiet sun (initial latitude and longitude of 37.2$\mydeg$ and 121.9$\mydeg$ with a tilt of -31.9$\mydeg$. hereafter quiet sun CMEs).  For both initial positions we consider all three background densities: ALF, TER1, and TER2.  Figure \ref{fig:HJdens} shows simulations for a 300 km s$^{-1}$, 10$^{14}$ g CMEs erupting from both initial locations for all three backgrounds (ALF in red, TER1 in white, and TER2 in blue).

For both initial locations we see that the CMEs deflect toward the Astrospheric Current Sheet.  In this case we see larger deflections from the quiet sun CMEs than the active region CMEs.  The active region CMEs initially deflect to the west until their longitudinal motion is halted and the CMEs begin a small latitudinal deflection.  The initial westward deflection results from imbalances in the local magnetic gradients in the active region.  This motion continues until the CMEs approach a region of enhanced magnetic field strength around 50$\mydeg$ longitude at 1.45$\rsun$.  This enhancement slows their longitudinal motion.  The ensuing latitudinal motion is a result of the global magnetic gradients determined by the location of the Astrospheric Current Sheet.  Much weaker local magnetic gradients deflect the quiet sun CMEs so their trajectory more closely resembles the direction of the global magnetic gradients.

For each initial position the CR2029 ALF and CR 2029 TER2 density models produce similar results, but the CR 2029 TER1 model yields a significant difference.  As the mass loss rate of the background stellar wind decreases the total amount of deflection increases due to the decrease in the drag, which results from the decrease in the stellar wind density.  Note that the \citet{Guh06} density model, used previously for solar ForeCAT simulations, produces results nearly identical to the ALF model.  Hereafter we only consider the CR2029 ALF and CR 2029 TER1 backgrounds and assume the CR2029 TER2 results do not differ from the CR2029 ALF results.

\begin{figure}[!hbtp]
\includegraphics[width=6.5in, angle=0]{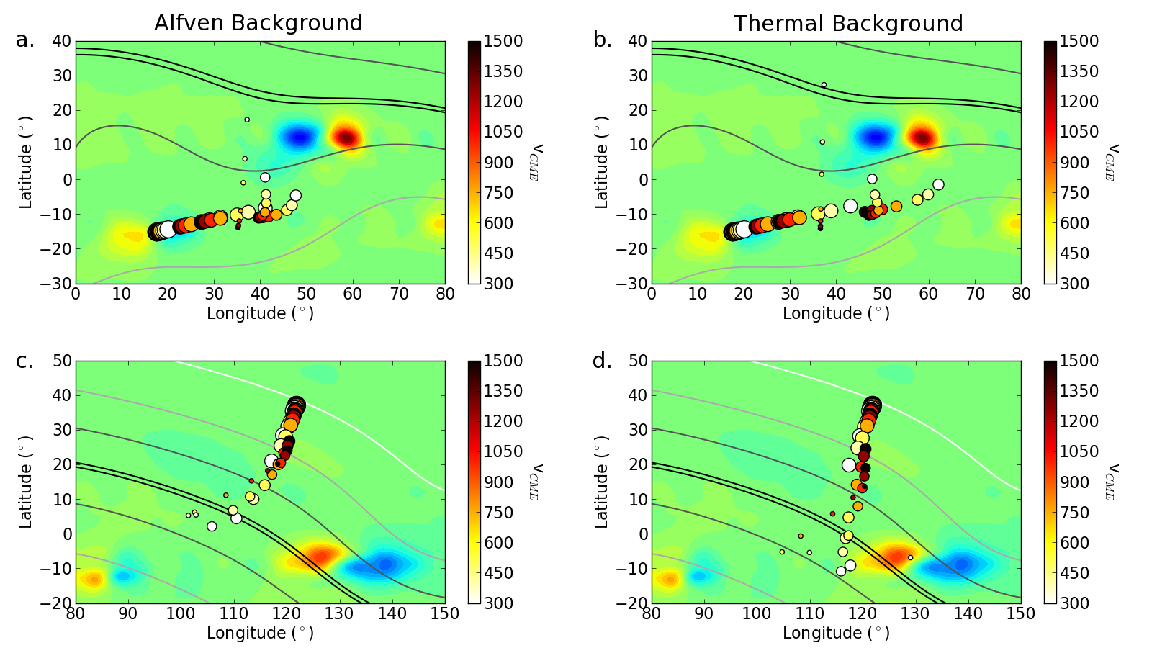}
\caption{Deflected positions of CMEs sampling a range of CME mass and speed for the solar-like star, analogous to Figure \ref{fig:MassPos}.  The left and right panels correspond respectively to results using the CR2029 ALF and CR 2029 TER1 background density.  The top and bottom panels show results for two different initial locations. }\label{fig:HJMvV}
\end{figure}

Figure \ref{fig:HJMvV} shows the deflected positions at 10 $\rsun$ for CMEs sampling a range of solar CME mass and velocities for the solar-type star, analogous to Figure \ref{fig:MvV}.  We include results for both initial positions using the CR2029 ALF and CR 2029 TER1 density models.  For the active region CMEs (top panels) we find that the smallest masses behave similar to the cases in Figure \ref{fig:HJdens} - they deflect westward until the region of enhanced magnetic field slows their longitudinal motion and the global gradients create a latitudinal motion.  The more massive CMEs deflect slower causing them to reach the enhanced magnetic field at farther distances where the enhancement has weakened and is unable to halt the longitudinal motion.  As we have seen for the M dwarf (and solar case in \citet{Kay15}) the deflection brings the CMEs closer to the Astrospheric Current Sheet and the amount increases with decreasing CME mass and speed.  This effect is more visible in the quiet sun CMEs (bottom panels) where the local gradients are relatively weak as compared to near the active region.  The behavior of these CME is the same as the solar CME we have previously studied \citep{Kay15}.  Since the majority of the deflection occurs below 10 $\rsun$ \citep{Kay15AM}, we do not expect significant difference between the deflected CME positions at hot Jupiter and solar habitable zone distances.

\subsection{Current Sheet Distance}
\begin{figure}[!hbtp]
\includegraphics[height=6in, angle=0]{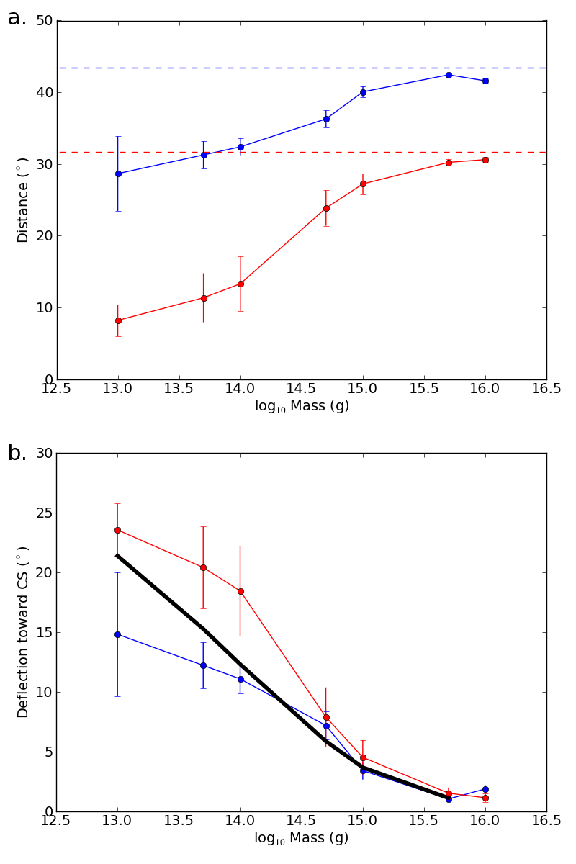}
\caption{Effects of deflection on the distance of the CMEs from the Astrospheric Current Sheet.  The top panel, analogous to Figure \ref{fig:CSdist} shows the distance from the Astrospheric Current Sheet, binned by mass for the two different initial locations.  The bottom panel shows deflection toward the Astrospheric Current Sheet, equivalent to difference between the final and initial distances.  The black line corresponds to the polynomial-best-fit to both cases.}\label{fig:HJCSdist}
\end{figure}

To quantify the amount of deflection we determine the average distance from the Astrospheric Current Sheet after the deflection as a function of CME mass, similar to Figure \ref{fig:CSdist}.  The results for the CR2029 ALF density model are shown in the top panel of Figure \ref{fig:HJCSdist}.  The blue line corresponds to the active region CMEs and the red line corresponds to the quiet sun CMEs.  The dashed lines show the initial distance of the CMEs.  As expected the distance from the current sheet decreases as the CME mass decreases, however we do not see a unique relation between mass and distance as seen for the M dwarf.  The M dwarf has much stronger forces that push the lowest mass CMEs entirely to the Astrospheric Current Sheet, the weaker solar forces can only cause motion toward the Astrospheric Current Sheet.   

Instead, for the solar case, it is more instructive to look at the deflection toward the Astrospheric Current Sheet as a function of CME mass.  We define the deflection toward the Astrospheric Current Sheet as the difference between the final distance, $\Delta_{ACS}$, and initial distance from the Astrospheric Current Sheet, $\Delta_{ACS,0}$.  The bottom panel of Figure \ref{fig:HJCSdist} shows the deflection toward the Astrospheric Current Sheet for both initial positions.  This quantity is better fit by a single function than the distance - the black line shows the polynomial best fit for masses below 5x10$^{15}$ g.
\begin{equation}\label{eq:deltaCSHJ}
\Delta_{\mathrm{ACS}} - \Delta_{\mathrm{ACS},0} = 1.007 \log\left(M_{\mathrm{CME}}\right)^3 -42.09 \log\left(M_{\mathrm{CME}}\right)^2 + 576.4 \log\left(M_{\mathrm{CME}}\right) - 2571
\end{equation}
We do not find a significant difference in the best fit polynomial between the CR2029 ALF and CR2029 TER1 results.
 
\section{Effect on Exoplanet CME Impacts}
From the ForeCAT CME deflections, we expect all CMEs to deflected toward the Astrospherical Current Sheet, but the magnitude of the deflection varies.  The less massive CMEs to deflect to and reach the Astrospheric Current Sheet for the M dwarf. More massive M dwarf CMEs and all stellar-type CMEs  deflect toward the Astrospheric Current Sheet, but do not necessarily reach it. Accordingly CME impacts should occur less frequently for planetary orbits inclined with respect to the Astrospheric Current Sheet.  Here we combine the results from ForeCAT with scaling laws from the literature to estimate the frequency of CME impacts as a function of a planet's orbital inclination, $i$, for both habitable zone M dwarf exoplanets and hot Jupiters orbiting a solar-type star.  In this work we refer to inclination from the plane of the Astrospheric Current Sheet, rather than with respect to the stellar equator, or relative to the plane of the sky.  Alternatively, this is equivalent to assuming the Astrospheric Current Sheet lies in the equatorial plane.

\subsection{Probability of Impact}
\citet{Kho07} use geometrical arguments to estimate the frequency of CME impacts for a planet with an equatorial orbit, assuming CMEs of angular width $\Delta$ are isotropically released between latitudes $\pm \Theta$.  \citet{Kho07} include an additional term, $\delta_P$, representing the planet's angular width but find it has a negligible effect so we do not include it.  The probability of impact, $P$, is calculated as 
\begin{equation}\label{eq:Kho07}
P = \frac{\Delta}{2\pi} \frac{\sin(\Delta/2)}{\sin(\Theta)}
\end{equation}
which is the product of a longitudinal and a latitudinal probability of the CME impacting the planet.

For orbital inclinations less than or equal to $\Theta$, the entirety of the orbit is contained in $\pm \Theta$, and there is no modification to the probability.  For inclinations greater than $\Theta$, we multiply Eq. \ref{eq:Kho07} by the fraction of the orbit between $\pm \Theta$.  This results in a probability of impact as a function of orbital inclination and CME latitude range.

\begin{equation}\label{eq:latprob}
P(i, \Theta) = \left\{
  \begin{array}{l l}
    \frac{\Delta}{2\pi} \frac{\sin(\Delta/2)}{\sin(\Theta)} & \quad i \le \Theta \\
    \frac{\Delta}{2\pi} \frac{\sin(\Delta/2)}{\sin(\Theta)}\left(1 - \frac{2}{\pi} \arccos(\frac{\sin(\Theta)}{\sin(i)})\right) & \quad i > \Theta
  \end{array} \right.
\end{equation}

\begin{figure}[!hbtp]
\includegraphics[height=8in, angle=0]{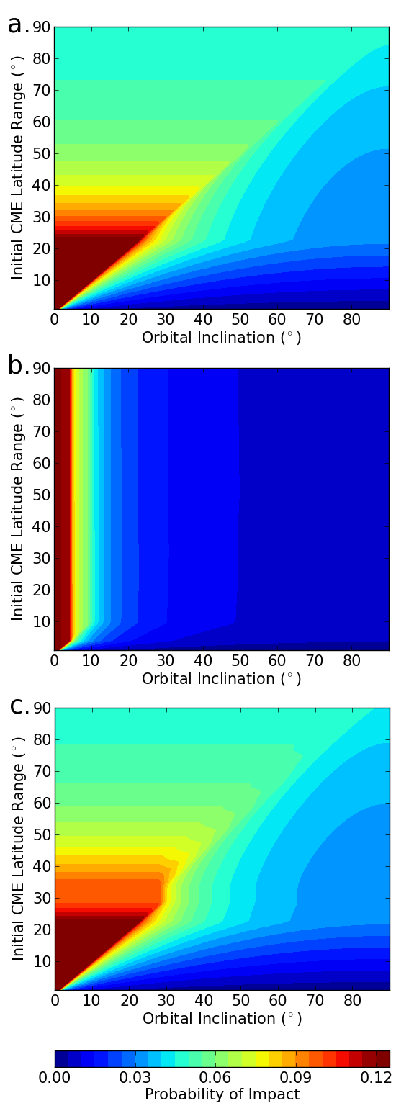}
\caption{Panel (a) shows the probability of CME impact versus orbital inclination with respect to the current sheet (Eq. \ref{eq:latprob}) for different values of the CME latitude range, $\Theta$.  Panels (b) and (c) show the change in the probability when the effects of deflections are included for a habitable zone M dwarf planet and a solar hot Jupiter.}\label{fig:latprob}
\end{figure}

We expect CMEs to erupt over a wide range of latitudes but the deflections will effectively collimate the CMEs into a smaller range of latitudes about the current sheet.  Figure \ref{fig:latprob}(a) shows Equation \ref{fig:latprob} for different values of $i$ and $\Theta$.  $\Delta$ is set to 60$\mydeg$ as in \citet{Kho07}.

Fig. \ref{fig:latprob}(a) shows that when $i \leq \Theta$ the probability is uniform as the orbit is fully contained within $\pm \Theta$.  If $i > \Theta$ the probability decreases as $i$ increases.  As $\Theta$ decreases, $P(i \leq \Theta)$ increases and $P(i>\Theta)$ decreases.  

\subsection{Incorporating CME Deflection}
We use Equations \ref{eq:deltaCS} and \ref{eq:deltaCSHJ} to incorporate the effects of deflection and determine the actual probability of impact.  Every CME mass has a different probability of planetary impact and the net probability is the sum of the individual probabilities weighted by the probability, $f$, of each mass.  We use discrete mass bins and approximate the post-deflection CME range $\Theta$ as equivalent to the post-deflection distance from the Astrospheric Current Sheet $\Delta_{CS}$.  For the M dwarf, a specific value of $\Delta_{ACS}$ is determined from Equation \ref{eq:deltaCS}.  
\begin{equation}\label{eq:Ptot}
P_{tot} = \sum\limits_l P(i, \Delta_{ACS}(M_l)) \; f(M_l)
\end{equation}
For the hot Jupiter we subtract the amount given by Equation \ref{eq:deltaCSHJ} from the initial CME distance from the Astrospheric Current Sheet.  A minimum distance of 0.5$\mydeg$, the smallest value from the M dwarf values, is assumed for the hot Jupiter results.

Equation \ref{eq:Ptot} requires the probability of a CME having a specific mass.  Using the observed M dwarf flare rates, the relationship between flare energy and CME mass, and the distribution of CME masses, and the probability of impact versus inclination, we can estimate the distribution of M dwarf CME masses.  For the solar-type CMEs we assume the solar distribution from \citet{Vou10}.

As discussed in Section \ref{dMparams}, \citet{Aud00} estimate approximately five 10$^{32}$ erg flares per day which corresponds to five ~10$^{17}$ g CMEs per day using the flare-CME mass relation from \citet{Aar12}.  \citet{Vou10} determine the probability $f(M)$ of a CME having mass, $M$. 
\begin{equation}\label{eq:PM}
f(M) = \frac{1}{\sigma \sqrt{2\pi}}\exp{\left(-\left(\frac{\ln(M) - \mu}{\sqrt{2} \sigma}\right)^2\right)}
\end{equation}
Using LASCO observations of 7668 solar CMEs, \citet{Vou10} determine an average mass, $e^\mu$, of 1.55x10$^{15}$ g, and a standard deviation, $\sigma$, of 1.114.  In section \ref{dMparams}, we estimate an average M dwarf CME mass of 1.66x10$^{16}$ g which we set as our new $\mu$, causing a shift in Eq. \ref{eq:PM} toward higher masses.  Using this new distribution, we determine a probability of 0.097 for a CME with mass 10$^{17}$ g.  If there are five 10$^{17}$ g CMEs then we expect a total number of CMEs per day, $N_{CME}$, of 51.  Using similar scaling relations and observations of V374 Peg flare rates, \citet{Vid16} estimate 15-60 CMEs per day with mass greater than 10$^{16}$ g.  These rates are about an order of magnitude higher than solar values during solar maximum, but this does not seem completely unreasonable given the increased activity of M dwarfs.  

Panel (b) of Figure \ref{fig:latprob} shows the values of Equation \ref{eq:Ptot} versus the initial CME latitude ranges and orbital inclination for the M dwarf.  Compared to the probability without deflection (Figure \ref{fig:latprob}(a)), the deflection causes the probability to increase for small inclinations and decrease for large inclinations.  For large initial CME latitude ranges, the probability roughly doubles for low inclinations, and decreases by a factor of 5 for high inclinations.  While subtle variations exist different initial CME latitude ranges, all have a probability of impact near 10\% for low inclinations and 1\% for high inclinations.  This corresponds to approximately 5 CME impacts per day for low inclinations and one impact every two days for high inclinations.  For comparison, roughly 10\% of solar CMEs are halo CMEs, half of which will propagate toward Earth opposed to away from Earth \citep{Web12}.  An average of 5 CMEs occur per day during solar maximum, which corresponds to an impact frequency of 0.25 CMEs per day.  Planets in an equatorial orbit of an habitable mid-type M dwarf may be impacted by CMEs 20 times more frequently than the average solar maximum rates at Earth.

Figure \ref{fig:latprob}(c) shows the probabilities for hot Jupiters orbiting a solar-type star when the effects of CME deflections are considered.  Comparison with panel (a) shows that deflections have less of an effect for the hot Jupiters than the mid-type M dwarf planets.  The probability increases slightly for low orbital inclinations and decreases slightly for high orbital inclinations.  Assuming solar-like CME rates, we expect between 0.05 and 0.5 CME impacts per day.  CMEs are certainly not isotropically released from the Sun, but they tend to occur below 50$\mydeg$ (polar crown filaments being a notable exception).  For an equatorial orbit and an initial latitude range of 50$\mydeg$, and assuming no changes between 10 $\rsun$ and 1 AU (which we expect to be true), we would determine an expected CME impact rate of 0.35 CMEs per day during solar maximum.  This agrees well with the number estimated from the fraction of halo CMEs.

We have not accounted for the short orbital period of hot Jupiters, which can be as short as 0.4 days \citep{Sah06}.  The above analysis gives a probability of impact that does not depend on the specific longitude of planet - all longitudes have equal probability.  A rapidly orbiting planet, however, will cover many different longitudes in the time it takes the CME to traverse the planet's orbital distance, increasing the chance of CME impact. 

\section{Magnetospheric Impacts}
So far we have shown that CME deflection can increase the number of CME impacts for both mid-type M dwarf exoplanets, and to a lesser extent, hot Jupiters.  In this section we determine what effect these impacts may have on a planetary magnetosphere.  \citet{Lam07} use a thermal balance model to determine the atmospheric profile of an exoplanet, which they then combine with a model for atmospheric pick-up ion loss driven by the external plasma conditions when a CME impacts the exoplanet.  They find significant atmospheric losses for magnetospheres that extend less than one planetary radii above the planet's surface.  Here, using a simple analytic model, we determine the minimum planetary magnetic field, $B_{p,min}$, required to maintain a magnetosphere of radius $r_m$ greater than 2 $r_p$, and we compare these values with the expected magnetic field strengths of rocky and gaseous planets. 

As in many previous works (\citet{Cha30}, \citet{Kiv95}, \citet{Vid13}, and references within), we balance pressure between the magnetosphere and the stellar wind or CME.  We assume the magnetospheric pressure is dominated by magnetic pressure and the external stellar pressure comes from both magnetic and ram pressure.  This yields
\begin{equation}\label{eq:Pbal}
\frac{[B_p (r_m)]^2}{8 \pi} = \frac{[B_{SW/CME}]^2}{8 \pi} + \frac{1}{2} \rho_{SW/CME} \; v^2_{SW/CME}
\end{equation}
where $B_p (r_m)$ is the magnetospheric field strength at a radial distance $r_m$ away from the planet's center, and $B_{SW/CME}$, $\rho_{SW/CME}$, and $v^2_{SW/CME}$ are the magnetic field strength, density, and speed of the stellar wind or CME upon reaching the planet.  Note that the magnetic field strengths in Equation \ref{eq:Pbal} corresponds only to the component of the magnetic field vectors tangential to the magnetopause.  For simplicity we consider the equator and assume the magnetopause normal parallels the radial direction. As in \citet{Vid13}, we model the planetary magnetosphere as a dipole:
\begin{equation}
B_p = B_{p,0} \left( \frac{r_p}{r} \right)^3
\end{equation}
where $B_{p,0}$ is the magnetic field strength on the surface at the equator.  At the equator $B_p$ has no radial component. For the stellar wind or CME magnetic field strength we consider only the tangential components of the magnetic field (toroidal and poloidal), which we describe below.

\subsection{M Dwarf Exoplanets}
We first consider the M dwarf case.  Our V374 ALF stellar wind model yields a density of 1.54x10$^{-17}$ g cm$^{-3}$ and a speed of 200 km s$^{-1}$ at the habitable zone distance of 0.1 AU.  The V374 TER model yields slightly different values with a density of 3.83x10$^{-18}$ g cm$^{-3}$ and a speed of 500 km s$^{-1}$ at the same distance.  The V374 ALF and V374 TER models produce ram pressures of 3.0x10$^{-3}$ dyn cm$^{-2}$ and 4.9x10$^{-3}$ dyn cm$^{-2}$, respectively.  Above the source surface, where only open magnetic field exists, the stellar magnetic field, $B_{SW}$, as a function of radial distance, $R$, can be determined from the Parker interplanetary magnetic field \citep{Par58}
\begin{equation}
\vec{B}_{SW} = B_{SS} \left(\frac{R_{SS}}{R}\right)^2 \hat{r} - B_{SS} \left(\frac{R_{SS}}{R}\right)^2 (R-R_{SS})  \frac{\Omega \sin\theta}{ v_{SW}}\hat{\phi}
\end{equation}
where $\hat{r}$ and $\hat{\phi}$ are the radial and toroidal directions.  $\theta$ is the colatitude, $B_{SS}$ is the magnetic field at the source surface distance, $R_{SS}$, and $\Omega$ is the stellar rotation rate.  We use the magnitude of the $\hat{\phi}$ component of  $\vec{B}_{SW}$ in Equation \ref{eq:Pbal}.  Previous pressure balance studies, such as \citet{Vid13}, have only considered the radial component of the interplanetary magnetic field, which, in its unperturbed state, is typically not tangential to the magnetopause.  It can, however, be argued that the draping of the interplanetary magnetic field about the magnetosphere will convert the radial component into a tangential component.   Previous studies have also typically neglected the toroidal component of the interplanetary magnetic field, which can become important in the case of fast rotators.  Additionally, any presence of a bow shock will affect the interplanetary magnetic field strength seen at the magnetopause.  Here we consider only the unperturbed tangential magnetic field and determine the minimum magnetic field stregth needed to prevent atmospheric losses due to the stellar wind.  Because $B_{SW}$ depends on the stellar wind speed, we obtain two different values for the two different background models: 1.9 G for V374 ALF, and 0.74 G for V374 TER, which correspond to magnetic pressures of 0.144 dyn cm$^{-2}$ and 0.0218 dyn cm$^{-2}$.  We determine $B_{p,min}$ by setting $r$ = 2 $r_p$.  We find minimum magnetic field intensities of 15 G and 6.6 G for V374 ALF and V374 TER, both roughly an order of magnitude larger than the Earth's magnetic field of ~1 G \citep{Bag92} and twice that of Jupiter \citep{Rus93}. 

Instead of stellar wind impact  we now consider the effects of the ram pressure and magnetic pressure of the CME itself.  CME-driven shocks, which we do not consider, should compress the upstream stellar wind, enhancing the external pressure upon the magnetosphere, and increasing the minimum planetary magnetic field.  We estimate the CMEs' magnetic field as it is not explicitly included in ForeCAT.  We adopt initial CME magnetic field strengths of either 1 kG or 20 kG, corresponding to the stellar quiet region or active region value.  Ignoring any reconnection, magnetic flux is conserved, so the magnetic field strength will decrease proportional to the increase in the cross-sectional area perpendicular to the magnetic field direction.  It can be shown for a self-similarly expanding toroidal CME and both poloidal or toroidal magnetic field that the cross-sectional area increases as $R^{-2}$.  For CMEs with 1 kG or 20 kG initial magnetic fields, we expect a magnetic pressure of 0.0088 dyn cm$^{-2}$ or 3.5 dyn cm$^{-2}$ at 20 $\rsun$.  Assuming self-similar expansion, our toroidal CMEs will have a volume proportional to $R^3$.  

Given the initial CME mass, which we approximate as constant, and volume, which we know from the CME shape, we can determine the density at any distance.  The ram pressure will vary greatly over the range of CME masses and velocities we consider, between 4.6x10$^{-6}$ dyn cm$^{-2}$ and 5100 dyn cm$^{-2}$.  For comparison, the average solar wind pressure at 1 AU is of order a few 10$^{-7}$ dyn cm$^{-2}$, and observations of Earth-impacting CMEs reach dynamic pressures as high as 2x10$^{-6}$ dyn cm$^{-2}$ (e. g. \citet{Cho10} and \citet{Lug15}).  The 08 March 15 CME had a dynamic pressure of 1.5x10$^{-6}$ dyn cm$^{-2}$, and the 2003 Halloween CME had a dynamic pressure of 7x10$^{-7}$ dyn cm$^{-2}$ at Mars and magnetic pressures of 1.6x10$^{-9}$ dyn cm$^{-2}$ and 4.3x10$^{-9}$ dyn cm$^{-2}$ \citep{Jak15}.  This implies that mid-type M dwarf exoplanets will need magnetic fields much stronger than found in our own solar system just to withstand the combined magnetic and ram pressure of the CMEs. 

\begin{figure}[!hbtp]
\includegraphics[width=6.5in, angle=0]{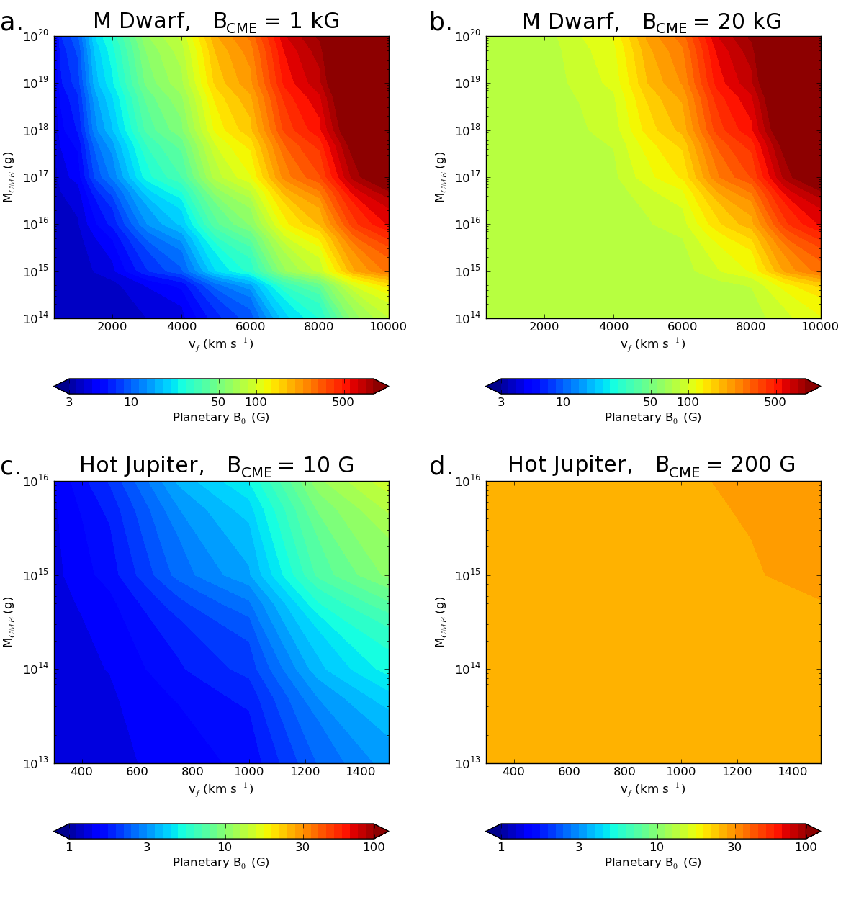}
\caption{Minimum planetary magnetic field strength required to sustain a magnetosphere twice the size of the planetary radius for different CME masses and speeds.  The top shows results for an M dwarf exoplanet, and the bottom shows results for a hot Jupiter.  The left and right columns show results for different initial CME magnetic field strengths.}\label{fig:planetB}
\end{figure}

Figure \ref{fig:planetB} shows $B_{p,min}$ for different CME masses and velocities.  The top row corresponds to mid-type M dwarf cases, and the left and right panels correspond to CMEs with initial magnetic field strengths of 1 kG and 20 kG.  Note the logarithmic scale color bar.  The faster, more massive CMEs require stronger planetary magnetic fields.  For these CMEs the ram pressure greatly exceeds the magnetic pressure.  Below CME masses of $~$10$^{15}$ g and $~$10$^{17}$ g the magnetic pressure exceeds the ram pressure for initial magnetic field strengths of 1 kG and 20 kG.

For CMEs with an initial magnetic field strength of 1 kG, we find that a planetary magnetic field of a few tens of Gauss can withstand a significant fraction of these CMEs.  \citet{Lop11} use the scaling law of \citet{Ols06} to estimate the dipolar moment of exoplanets.  For fast rotators, \citet{Lop11} predict dipole moments up to 80 times larger than Earths, and determine dipole moment up to 18 times Earth's for specific observed exoplanets.  It seems unlikely, however, for a rocky body to generate a magnetic field of order hundreds to thousands of Gauss, so the fastest, most massive CMEs are capable of compressing the planetary magnetosphere below 2 $r_p$.  These CMEs, however, are the ones that are the least likely to impact a planet.  Deflection forces cannot significantly influence their motion so there is no enhancement in the impact probability.  

Figure \ref{fig:planetB}(b) shows that when the initial magnetic field strength is increased to 20 kG only planets with magnetic field strengths larger than 100 G will be shielded from the majority of CMEs.  This includes the slowest, least massive CMEs, which do have an enhanced probability of impacting the planet.  We conclude that knowledge of the magnetic field of M dwarf CMEs is critical for understanding whether a planet's atmosphere could possibly withstand continual impacts for any extended period of time.

\subsection{Hot Jupiters}
We repeat the analysis in the previous section for a hot Jupiter orbiting a solar-type star at 10 $\rsun$.  As compared to the M dwarf exoplanet, the hot Jupiter experiences weaker ram and magnetic pressure from the solar-like wind.  For the CR2029 ALF background, we find a magnetic pressure of 1.53x10$^{-8}$ dyn cm$^{-2}$, and a ram pressure of 1.87x10$^{-6}$, which corresponds to a minimum planetary magnetic field of 0.055 G.  The CR2029 TER2 background does not differ significantly from the CR2029 ALF background.  For the less dense CR2029 TER background we determine a magnetic pressure of 1.03x10$^{-8}$ dyn cm$^{-2}$, and a ram pressure of 1.22x10$^{-6}$ dyn cm$^{-2}$, which corresponds to a magnetic field strength of 0.044 G.  For all backgrounds we find a minimum planetary magnetic field strength significantly less than found for the M dwarf exoplanet, and less than that of the gaseous planets within our own solar system.

We compare the minimum planetary magnetic field strength needed to withstand the stellar wind with the magnitude needed to withstand the ram and magnetic pressure experienced during a CME impact.  Again we determine values for the range of CME masses and speeds considered in the analysis of CME deflections.  This range corresponds to ram pressures between 5.88x10$^{-6}$ dyn cm$^{-2}$ and 0.14 dyn cm$^{-2}$.  We consider initial CME magnetic field strengths of 10 G and 200 G, which corresponds to the magnitude of the solar quiet sun and active regions in the photosphere.  This yields magnetic pressures of 0.001 dyn cm$^{-2}$ and 0.45 dyn cm$^{-2}$.

Figure \ref{fig:planetB}(c) and (d) show contours of the minimum planetary magnetic field for the exoplanet cases.  For the lower magnetic field CME (Figure \ref{fig:planetB}(c)), we find that the ram pressure dominates for most of the CMEs, leading to significant variation across parameter space.  The least massive, slowest CMEs require a planetary magnetic field of 1.32 G, and this increases to 15.5 for the most massive, fastest CMEs.  For the stronger magnetic field CME, the magnetic pressure dominates leading to a larger minimum planetary magnetic field, and less variation with CME parameters.  In this case the minimum planetary magnetic field varies between 26.9 G and 30.9 G. 

While these magnetic field strengths may slightly exceed those within our own solar system, they may not be unreasonable for hot Jupiters.  \citet{Chr09} extends a scaling model derived in \citet{Chr06}, which relates the energy flux available to drive a dynamo with the resulting magnetic field strength.  This model reproduces the magnetic field strength of the Earth and Jupiter, as well as some low-mass stars, and predicts a magnetic field strength 10 times that of Jupiter for a planet 10 times as massive as Jupiter.  This suggests that Jupiter-like planets could have a magnetic field of order 50 G.  Recent work by \citet{Cau15} infers a 28 G planetary magnetic field based on the standoff distance of the planetary bow shock.  This suggests that hot Jupiters may have magnetospheres that can shield them from the effects of CME impacts.

\section{Discussion}\label{disc}
While CME deflection will cause a small increase in the frequency of CME impact for a hot Jupiter orbiting a solar-type star, and the hot Jupiter may need a slightly stronger magnetic field, the system is not terribly different from our own solar system.  In comparison, CMEs from a mid-type M dwarf present an extreme version of solar CMEs.  For a mid-type M dwarf habitable zone planet to be truly habitable requires planetary magnetic field strengths orders of magnitude greater than found in our own solar system.  In this work, however, we have only considered one M dwarf of spectral type M4, and at only one point in its main-sequence lifetime.  Activity tends to increase from early- to late-type M dwarfs \citep{Moh02, Wes15}, until above type M9 when activity decreases \citep{Moh03}.  Late-type M dwarfs tend to have the longest activity lifetimes \citep{Wes08}.  \citet{Rei12LR} show that young, rapidly-rotation, early-type M dwarfs can have kG magnetic fields, and these magnetic field strengths are common among mid- and late-type M dwarfs over a wider range of ages.  We expect CMEs to accumulate around the Astrospheric Current Sheet for young early-type M dwarfs and most mid- and late-type M dwarfs.  Depending on a star's activity lifetime, there may be some time after which the stellar magnetic field and/or activity weakens, constant CME bombardment ceases, and an previously inhabitable exoplanet can start developing an atmosphere.

The activity of M dwarfs tends to correlate with stellar rotation for early-type M dwarfs (e.g. \citet{Piz03}; \citet{Moh03}; and \citet{Kir07}).  \citet{Rei12} develop a model for the spin-down time of a low-mass star due to angular momentum loss via the stellar wind.  We suggest that constant, massive CMEs, could also have an effect on the angular momentum loss of a star, which in turn would affect the activity levels.  The analysis in section 7 suggests that as many as 50 CMEs per day may occur for an active M dwarf, which agrees with a recent estimation for V374 Peg by \citet{Vid16}.  While this order-of-magnitude estimate is extreme by solar standards, we still expect that less frequent CMEs could affect the angular momentum loss.  Better understanding will only occur with improved observations of inferred CMEs outside our own solar system.  Estimating CME's contribution also requires knowledge of the Alfv\'{e}n radius.  We can make an estimate from our simple static backgrounds, but we expect dynamic effects may strongly affect this value.  If the CME legs remain attached to the surface of a star, they will begin to wrap around it as the star rotates as much as several times a day.  Barring any reconnection, this could significantly enhance the magnetic field strength close to the star as it crease a large toroidal magnetic field.  This would increase the Alfv\'{e}n radius, allowing for more efficient magnetic breaking of the stellar rotation.

Similarly one might wonder the effect of the large numbers of CMEs on the stellar mass loss rate.  Using the M dwarf CME mass distribution used in this work, and assuming 50 CMEs a day yields a mass loss rate of 1.5x10$^{-13}$ M$_{\Sun}$ yr$^{-1}$.  M dwarfs' main sequence lifetimes exceed the solar main-sequence lifetime of 10 billion years, but we find that an M dwarfs mass would change by 0.0015 M$_{\Sun}$ yr$^{-1}$ over a solar lifetime.  While this is not a insignificant percentage of the total M dwarf mass, we do not expect mass loss from CMEs to cause an M dwarf to slowly change spectral type.  However, if the actual mass-loss rate happens to be much higher than our calculated value, it could start to become a concern.

Finally, throughout this work we have focused on how CME impacts may adversely affect an exoplanet.  In contrast, frequent CME impacts benefit astronomers searching for signatures of aurora on exoplanets.  Radio detections of extrasolar could identify previously undiscovered exoplanets, as well as information about exoplanetary magnetospheres.  Recently \citet{Hal15} found signatures of aurora on a planet orbiting an M8.5 dwarf.  Continued observations with existing, and new facilities such as the Owens Valley Long Wavelength Array (OV-LWA), should find more signatures of planetary magnetospheres.  We suggest the best candidates for extrasolar auroral detections are those where CME impacts occur the most frequently.  Exoplanets orbiting M dwarfs should experience the most frequent CME impacts, and we expect these CMEs to have a stronger magnetospheric impact than solar CMEs.  We have shown that hot Jupiters require a significant, yet not implausible, planetary magnetic field, which, combined with their larger size, would lead to aurora brighter than those in our own solar system. 

\section{Conclusion}
We have adapted ForeCAT, a solar CME deflection model, to simulate CME deflections for the M4 dwarf V374 Peg.  M dwarf CME parameters are highly uncertain so we consider CME masses between 10$^{14}$ g and 10$^{19}$ g and final propagation speeds between 300 km s$^{-1}$ and 10,000 km s$^{-1}$.  The deflection depends strongly on the CME mass, and, to a much weaker extent, the CME speed.  The least massive CMEs quickly deflect to the Astrospheric Current Sheet and remain trapped there. These extreme deflections exceed those seen in our own solar system.

We also apply ForeCAT to the deflections of CMEs for solar-type stars at hot Jupiter distances.  We find the same dependence on mass and speed as seen for the M dwarf but the magnitude of the deflections is significantly smaller.  While the CMEs continue to deflect toward the Astrospheric Current Sheet, only the slowest, least massive CMEs actually reach it by 10 $\rsun$, analogous to what we see in our own solar system.  

We determine the average distance between the CME's deflected position and the Astrospheric Current Sheet as a function of CME mass.  For the mid-type M dwarf CMEs we find that the distance does not depend on the initial position for CME masses below 10$^{17}$ g.  For the solar-like CMEs the distance does depend on the initial position so we determine the amount of deflection toward the Astrospheric Current Sheet as a function of CME mass.  We combine these relations between the deflection and the CME mass with geometrical arguments and an estimation of the M dwarf CME mass distribution to determine the probability and frequency of CME impacts.  

For both habitable zone mid-type M dwarf exoplanets and hot Jupiters the probability of impact decreases if the exoplanet's orbit is inclined with respect to the Astrospheric Current Sheet.  The sensitivity to the inclination is much greater for the mid-type M dwarf exoplanets due to the extreme deflections to the Astrospheric Current Sheet.  For low inclinations we find a probability of 10\% whereas the probability decreases to 1\% for high inclinations. From our estimation of 50 CMEs per day, we expect habitable mid-type M dwarf exoplanets to be impacted 0.5 to 5 times per day, 2 to 20 times the average at Earth during solar maximum.  The frequency of CME impacts may have significant implications for exoplanet habitability if the impacts compress the planetary magnetosphere leading to atmospheric erosion.  For the hot Jupiters, the impact probability has a similar range but more inclinations have moderate values, near a few percent, as opposed to a clear division between the two extremes as seen for the mid-type M dwarf.

We quantify the effect of these CMEs on an exoplanetary magnetosphere by determining the minimum planetary magnetic field need to maintain a magnetosphere twice the size of the planetary body.  This size corresponds to the minimum necessary to retain an atmosphere according to \citet{Lam07}.  We find the mid-type M dwarfs exoplanets require planetary magnetic fields between tens to hundreds of Gauss, whereas hot Jupiters only require magnitudes between a few and 30 G.  The magnetic field needs to shield a planet from CME impacts greatly exceeds that required for shielding from the stellar wind.  We expect that rocky exoplanets cannot generate sufficient magnetic field to shield their atmosphere from mid-type M dwarf CMEs, but hot Jupiters around solar-type stars could likely remain shielded.  We expect that the minimum magnetic field strength will change with M dwarf spectral type as the amount of stellar activity and stellar magnetic field strength change, and that early-type M dwarfs would be more likely to retain an atmosphere than mid or late-type M dwarfs.

\acknowledgements
M.O. would like to acknowledge the support NSF CAREER Grant ATM-0747654. 


\end{document}